\documentclass[manuscript]{aastex}
\usepackage{emulateapj5}
\usepackage{amsmath}

\shorttitle{The New PWN G141.2$+$5.0}
\shortauthors{R. Kothes et al.}

\begin{document}

\title{G141.2$+$5.0, a new Pulsar Wind Nebula discovered in the Cygnus Arm of the Milky Way}

\author{R. Kothes\altaffilmark{1}, 
  X.H. Sun\altaffilmark{2},
  W. Reich\altaffilmark{3},
  T.J. Foster\altaffilmark{4,1}
}

\altaffiltext{1}{National Research Council of Canada, 
              Herzberg Programs in Astronomy and Astrophysics,
              Dominion Radio Astrophysical Observatory,
              P.O. Box 248, Penticton, British Columbia, V2A 6J9, Canada}

\altaffiltext{2}{Sydney Institute for Astronomy, School of Physics, 
                 The University of Sydney, NSW 2006, Australia}

\altaffiltext{3}{Max-Planck-Institut f\"ur Radioastronomie, 
              Auf dem H\"ugel 69, 53121, Bonn, Germany}

\altaffiltext{4}{Department of Physics and Astronomy, Brandon University, 
             270 18th Street, Brandon, MB R7A 6A9, Canada}

\email{roland.kothes@nrc-cnrc.gc.ca}

\begin{abstract}
We report the discovery of the new pulsar wind nebula (PWN) G141.2$+$5.0 in 
data observed with the Dominion Radio Astrophysical Observatory's Synthesis 
Telescope at 1420~MHz. The new PWN has a diameter of 
about 3.5\arcmin, which translates at a distance of 4.0~kpc to a spatial extent 
of about 4~pc. It displays a radio spectral index of $\alpha\approx -0.7$, 
similar to the PWN 
G76.9$+$1.1. G141.2$+$5.0 is highly polarized up to 40\,\% with an average of 
15\,\% in the 1420 MHz data. It is located in the centre of a small spherical HI bubble, 
which is expanding at a velocity of 6~km\,s$^{-1}$. The bubble's systemic 
velocity is $-53$~km\,s$^{-1}$ and could be the result of the progenitor star's
mass loss or the shell-type SNR created by the same supernova 
explosion in a highly advanced stage. The systemic velocity of the bubble 
shares the velocity of \ion{H}{1} associated with the Cygnus spiral arm, which 
is seen across the 2nd and 3rd quadrants and an active star-forming arm
immediately beyond the Perseus arm. A kinematical distance of 4$\pm$0.5~kpc is 
found for G141.2$+$5.0, similar to the optical distance of the Cygnus arm 
(3.8$\pm$1.1~kpc). 
G141.2$+$5.0 represents the first radio PWN discovered in 17 years and the
first SNR discovered in the Cygnus spiral arm, which is in stark contrast
with the Perseus arm's overwhelming population of shell-type remnants. 

\end{abstract}


\section{Introduction}


Pulsar wind nebulae (PWNe) are created by fast spinning neutron stars that produce highly relativistic 
magnetized particle winds, inflating an expanding bubble,
confined by the expanding ejecta of the supernova explosion. Electrons and positrons are
accelerated to relativistic speeds at the termination shock and interact with the magnetic field
generating a synchrotron emitting nebula, called a pulsar wind nebula.

A PWN is characterized by a central concentration of radio continuum emission that gradually 
declines to the outer edge, in contrast to the more common shell-type supernova remnants (SNRs),
which have a limb-brightened appearance. PWNe typically have flat radio
continuum spectra with a spectral index $\alpha$
between 0.0 and $-$0.3 ($S\sim\nu^\alpha$, $S$: flux density, $\nu$: frequency), although steeper
spectra have been observed for PWNe DA495 \citep{koth08b} and G76.9$+$1.1 
\citep{land93,koth06}.
Many of these PWNe, in particular most of the very bright objects like the
Crab nebula and 3C58, do not show an accompanying shell-type SNR. This is very puzzling
since typically they are relatively young so that the actual remnant of the
supernova should not have dissipated, yet. 

The Canadian Galactic Plane Survey \citep[CGPS,][]{tayl03} has portrayed the
outer Galactic plane with excellent sensitivity and fidelity to extended
emission in radio continuum and HI. These qualities have led to the
discovery of 13 supernova remnants (SNRs) over the past 12 years
\citep{koth01b,koth03b,koth05,tian07,fost13,gerb13}. 

\section{Observations and Data Processing}

Radio continuum observations at 408~MHz and 1420~MHz were obtained with the 
Dominion Radio Astrophysical Observatory's Synthesis
Telescope \citep[ST,][]{land00} 
The main characteristics of this survey are the same
as for the CGPS \citet{tayl03}. 
 
The DRAO ST provides observations of linearly polarized emission at four frequency bands
around the HI line to allow precise determination of rotation
measures (RMs). Those bands are 7.5~MHz wide with center frequencies at 1406.9~MHz (band A),
1413.8~MHz (band B), 1427.4~MHz (band C), and 1434.3~MHz (band D).

Angular resolution varies slightly across the final maps as cosec(Declination). At the centre
of G141.2$+$5.0 the resolution in the final maps is $56\arcsec \times 48\arcsec$ (DEC $\times$ R.A.)
for 1420~MHz continuum, $68\arcsec \times 58\arcsec$ for HI, and 
$3\farcm2 \times 2\farcm8$ for 408~MHz continuum. The RMS noise in the final images is
180~$\mu$Jy/beam for the 1420~MHz continuum data, 2~K~T$_b$ for the HI data, and 2~mJy/beam for
the 408~MHz data.

\section{Results}

\subsection{G141.2+5.0 in Radio Continuum}

\begin{figure*}[!ht]
\begin{center}
  \includegraphics[bb = 240 45 490 705,height=18cm,angle=-90,clip]{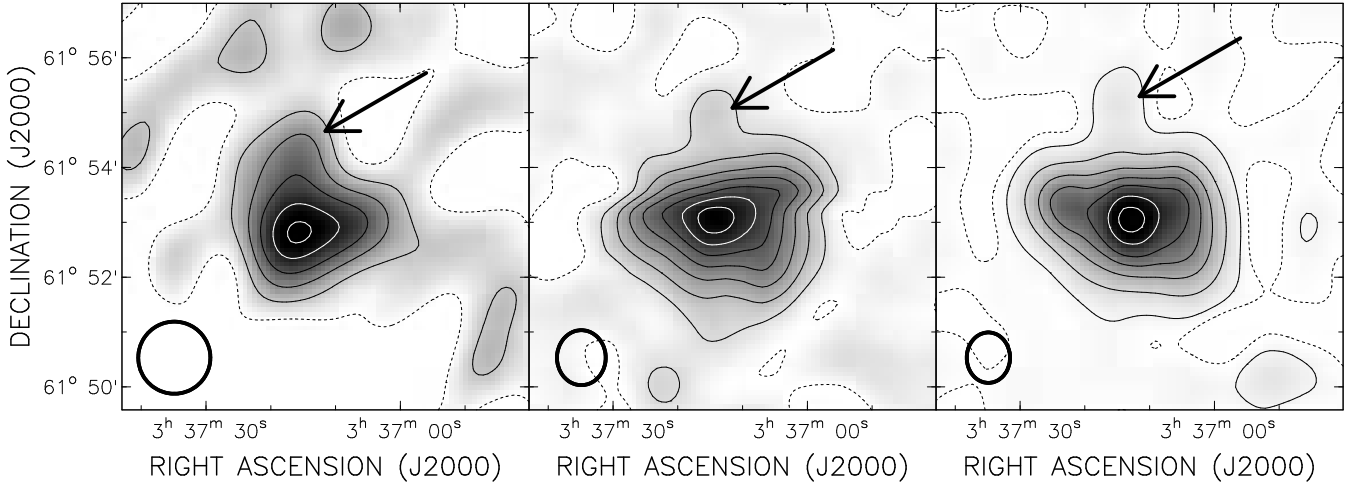}
\end{center}
 \caption{\label{fig:tp} Total Power images of the new PWN G141.2$+$5.0. From the left to the right:
VLSS 74 MHz, WENSS 327 MHz, and DRAO ST 1420 MHz. Contours are shown at 100, 200, 300, 400, and 470~mJy/beam 
at 74~MHz, at 10, 20, 30, 40, 50, 60, and 70~mJy/beam at 327~MHz, and at 1, 4, 8, 12, 16, 22, and 27~mJy/beam 
at 1420~MHz. The dashed contour indicates the zero level. The angular resolution of the observations is 
indicated by the circle in the lower left corner. The location of the unrelated point source is 
indicated by an arrow.}
\end{figure*}

The newly discovered extended radio continuum source G141.2+5.0 is shown in Figure~\ref{fig:tp} 
in total power. For comparison, we also display images from the
Westerbork Northern Sky Survey (WENSS) at 327~MHz \citep{reng97} and the VLA Low-Frequency Sky Survey Redux
(VLSSr) at 74~MHz \citep{lane12}. 
In the 1420~MHz image, with the best image quality and resolution, this source
appears as a diffuse extended radio source with a bright compact core. It is somewhat elongated along R.A.
and has a length of $3\farcm5$ and a height of $3\arcmin$. 

For the radio spectrum determination we use the catalogue flux density at 232~MHz from the
Miyun 232~MHz survey \citep{zhan97}, where G141.2$+$5.0 is not resolved. The flux density at 
5~GHz was determined with the SPECFIND cross identification catalogue \citep{voll10}. 
For the other surveys the flux density has been integrated in concentric rings starting at the radio peak of the 
source at 1420~MHz ($\mathrm{\alpha_{J2000}=3^h 37^m 12.6^s,\  \delta_{J2000}}=61\degr 53\arcmin 5\arcsec$). We
used this approach rather than simply using the catalogued fluxes, 
because sources are typically fitted with a Gaussian model, which may not describe a 
resolved source very well. 

In Figure~\ref{fig:tp} a small extension from the source to the north is apparent, most 
likely an unrelated, extragalactic steep-spectrum source. It is obvious from the images in
Figure~\ref{fig:tp} that this source becomes more dominant for longer wavelength. In the DRAO
ST and the WENSS images it is clearly a weak insignificant object, but in the VLSSr image it 
becomes a bright source with significant flux contribution. We fitted a Gaussian to this point
source in the 1420~MHz maps and the WENSS data to determine its spectrum (Fig.~\ref{fig:spec+rm}).
We removed its contribution from the integrated flux densities of the extended 
source.

The final flux densities are listed in Table~\ref{tab:flux} and the integrated radio continuum 
spectrum of G141.2+5.0 is displayed in Figure~\ref{fig:spec+rm}.
We found a straight spectrum between 74~MHz and 5~GHz with a spectral index of $\alpha \approx -0.7$.

\subsection{Polarization Properties of G141.2+5.0}

\begin{figure}[!ht]
\begin{center}
  \includegraphics[bb = 50 120 400 450,width=8cm]{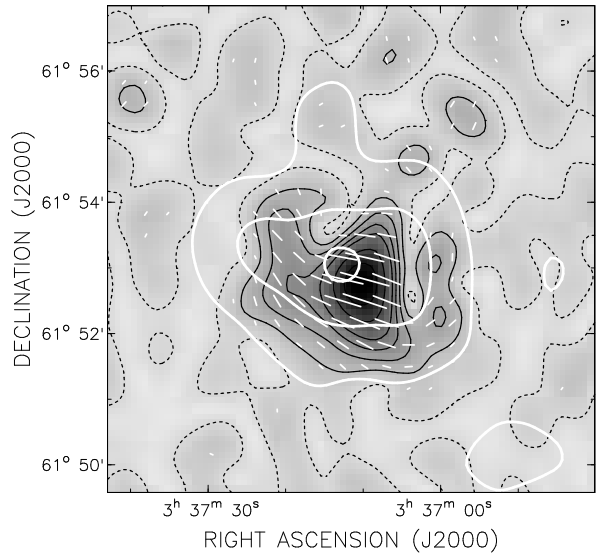}
\end{center}
 \caption{\label{fig:pi} Polarized intensity image of the new PWN G141.2$+$5.0. Black contours are shown at
0.4, 0.8, 1.2, 1.6, 2.4, 3.2, and 4.5~mJy/beam. The dashed contour indicates the zero level. White contours 
represent the total power emission. Polarisation vectors are overlaid in E-field direction.}
\end{figure}

\begin{figure}[!ht]
\begin{minipage}{9cm}
  \includegraphics[bb = 80 200 525 600,width=9cm]{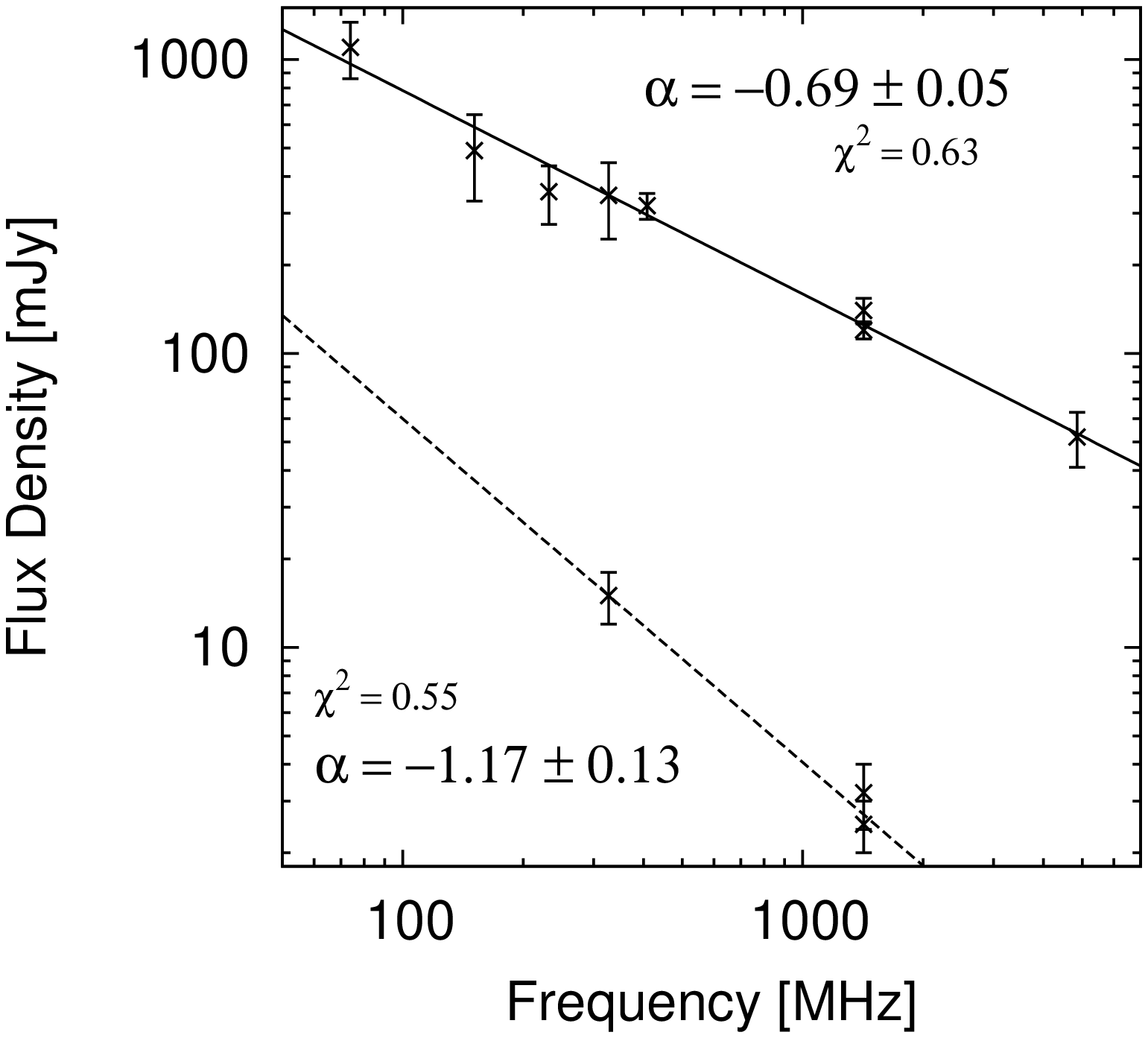}
\end{minipage}
\hfill
\begin{minipage}{8cm}
  \includegraphics[bb = 80 175 525 620,width=8cm]{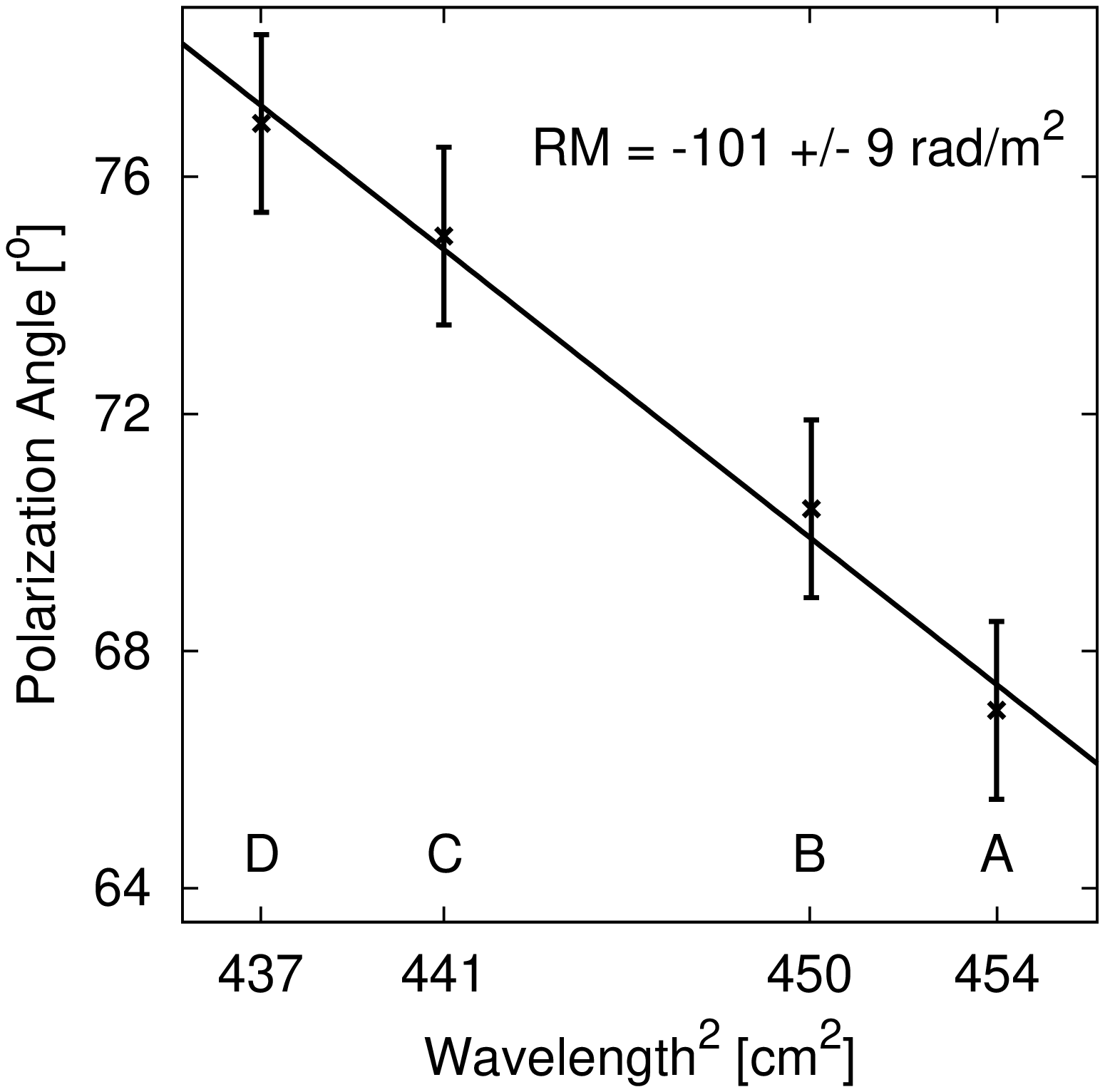}
\end{minipage}
 \caption{\label{fig:spec+rm} Left: Radio continuum spectrum of the new PWN G141.2$+$5.0 (solid) and the
 nearby point source (dashed). The spectral indeces and reduced $\chi^2$ are indicated. Right:
 RM spectrum of the highly polarized core of G141.2$+$5.0, calculated with the four bands of the DRAO
 ST observations. The four bands of the DRAO ST observations are indicated.}
\end{figure}

The polarization image of G141.2+5.0 is shown in Figure~\ref{fig:pi}. 
In polarization, similar to total power, it has a compact bright core, 
but its peak is offset from the total power peak and it is
elongated north-south. There are two faint arc-like features on either side of the core visible in
polarized intensity.

The integrated polarized flux in the DRAO ST observations is 
$S_{PI} = 18 \pm 3$~mJy. This results in a total fractional polarization of 15~\% at 1420~MHz.
The peak polarization is more than 40~\%.

Since the DRAO ST provides observations of linearly polarized emission at four frequency bands
we are able to calculate RMs. In our observations of G141.2+5.0
only the polarized core is sufficiently bright to have a large enough signal-to-noise 
ratio in the individual bands. 
In Figure~\ref{fig:spec+rm}, we provide a RM spectrum for this central core. The resulting Faraday rotation
is $RM = -101 \pm 9$~rad\,m$^{-2}$.
G141.2+5.0 is listed in the NVSS rotation measure catalogue \citep{tayl09} with an
RM of $-37.8 \pm 16.0$~rad\,m$^{-2}$. However, \citet{tayl09} just took the RM at the
location of the total power peak of each source. In the case of G141.2+5.0, which is an extended
source in the NVSS, there
is a strong gradient in polarized intensity (see Figure~\ref{fig:pi}), making 
this value unreliable.

From our data alone we cannot distinguish between foreground and internal Faraday rotation. The
3D-models of the Galactic magnetic fields, cosmic-ray, and thermal electrons by
\citet{sun08} and \citet{sun10} give a foreground rotation measure of $-25$~rad\,m$^{-2}$ for the direction of
G141.2+5.0 and a distance of 4~kpc (see section 4.2), consistent with values found in the NVSS RM 
catalogue \citep{tayl09} in this area of the Galaxy. This indicates significant internal Faraday rotation
for G141.2+5.0.

\subsection{The Environment of G141.2+5.0}

\begin{figure}[!ht]
\begin{center}
  \includegraphics[bb = 30 30 535 585,height=15cm,angle=-90,clip]{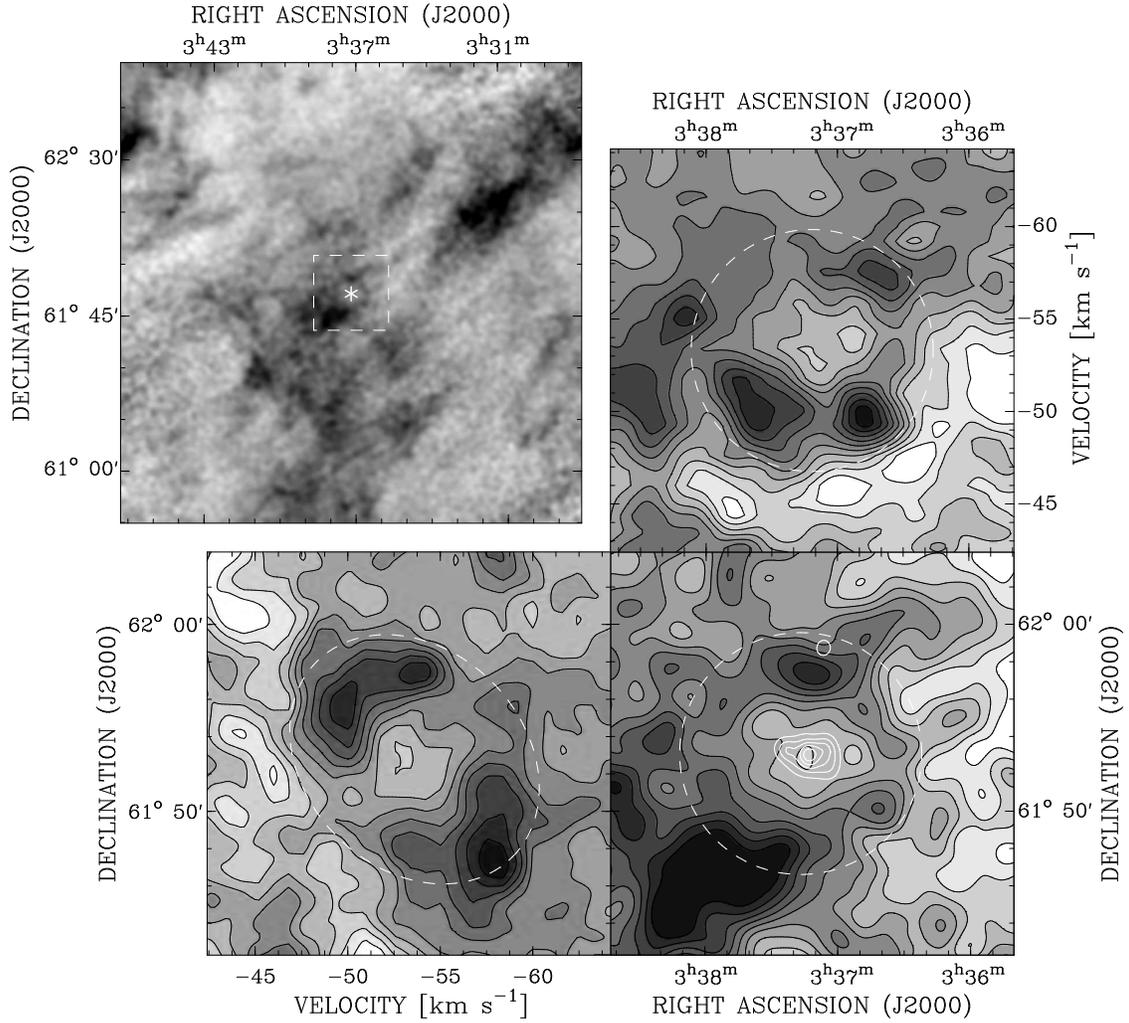}
\end{center}
 \caption{\label{fig:hi} HI maps of the area around G141.2$+$5.0. All images have been convolved to a resolution
 of 2\arcmin.
 Top left: HI channel map at a radial velocity of about $-53$~km\,s$^{-1}$ with single antenna data added taken from the 
 Low-Resolution-DRAO-Survey \citep[LRDS][]{higg00} observed with the DRAO 26~m single antenna telescope. 
 The gray-scale goes from a brightness temperature of 10~K (white) to 40~K (black). The location of G141.2$+$5.0
 is marked by the asterisk. The white box indicates the angular extent of the other three maps. 
 Bottom right: HI channel map at a radial velocity of $-53$~km\,s$^{-1}$ zoomed in as indicated by the white box
 in the top left panel. The radio continuum emission of G141.2$+$5.0 is indicated by the white contours. Top right:
 R.A.-velocity representation of the image in the bottom right. Bottom left: Velocity-DEC representation of the
 image in the bottom right. In all images the dashed 
 circle outlines the putative expanding HI shell.}
\end{figure}

In Figure~\ref{fig:hi}, we display HI images of the area around G141.2+5.0. In the top left panel a large area
around G141.2+5.0 is displayed. 
In the centre a small, narrow ring-like feature is
discernable surrounding G141.2+5.0. 

In the lower right panel, we zoomed in on this ring-like feature. G141.2+5.0 
is located in the centre of a depression, which is in most directions surrounded by
a broken HI shell. This HI shell is also visible in the R.A.-v and v-DEC slices displayed 
in Figure~\ref{fig:hi} 
as a distinct feature. The distribution of the HI emission indicates that this HI feature
actually is a broken HI shell expanding at a rate of $6\pm 1$~km\,s$^{-1}$ as indicated by the 
dashed white circles. This is probably an 
old HI bubble that has slowed down to a velocity close to the sound speed of the ambient 
medium and is therefore breaking up.
The systemic velocity can be derived from the circles indicating the extend
of the HI structure in the R.A.-v and v-DEC slices to be
$-53\pm 1$~km\,s$^{-1}$.

\section{Discussion}

In an ongoing survey of an area in the so-called Fan Region, we have serendipitously discovered a
previously unidentified diffuse and extended ($\gtrsim 3\arcmin$) radio continuum source
G141.2+5.0. The Fan 
region is a highly linearly polarized feature that
extends above the mid-plane from a Galactic Longitude of $130\degr$ to just beyond the 
anti-centre \citep[see Figure 11 in][]{woll06}. G141.2+5.0
is highly linearly polarized with an integrated fractional
polarization of about 15~\% and a peak of more than 40~\%. It shows a steep radio continuum
spectrum with a spectral index of $\alpha \approx -0.7$. 


\subsection{The nature of G141.2+5.0}

We have tried to identify this source in surveys available through the SkyView Virtual Telescope
Database\footnote{http://skyview.gsfc.nasa.gov/cgi-bin/titlepage.pl}. But we found no counterpart outside 
the long-wavelength radio regime. Since, in particular, we did not
find any counterpart in the infrared we can dismiss the possibility that this source may be an HII region
or a nearby Galaxy. 

One possible explanation we like to briefly discuss is diffuse synchrotron emission related to a Galaxy cluster, 
like a halo or a relic, even though the HI shell we found seems to locate G141.2+5.0 inside the Milky
Way Galaxy. 

\citet{ferr08} report that radio haloes are filling the central areas of Galaxy clusters. They have 
diffuse quite regular structure with very low linear polarization, low radio surface brightness, 
and very steep radio continuum spectra with $\alpha \lesssim -1.0$. The low fractional polarization
and the very steep radio spectra typically found in haloes make this an unlikely
explanation for G141.2+5.0. G141.2+5.0 is too bright, highly polarized, and the radio
spectrum is too flat.

\citet{ferr08} also discuss that cluster radio relics are generally filamentary features and 
commonly found in merging clusters mostly related to shocks. They can be highly linearly polarized 
at the 10 - 30\,\% level and are located at the cluster periphery. Similar to the haloes they
have steep radio continuum spectra with $\alpha \lesssim -1.0$. Even though some characteristics
of relics are similar to those we found in G141.2+5.0, the steep radio spectra and the filamentary
structure make it very unlikely that G141.2+5.0 is a radio relic of a Galaxy cluster. 

In particular, since the extended radio source G141.2+5.0 seems to be located inside an expanding
HI bubble, but also because of the characteristics of its radio continuum emission,
the remnant of a supernova is the most likely explanation. While the structure
of G141.2+5.0, a central concentration of radio continuum emission that gradually 
declines to the outer edge, indicates a pulsar wind nebula, the steep
radio continuum spectrum is more reminiscent of a shell-type remnant. However, there are
examples for steep spectrum pulsar wind nebulae. The PWN DA\,495 has a much steeper radio
continuum spectrum than G141.2+5.0 beyond its cooling break frequency of 1.3~GHz \citep{koth08b}
and the PWN G76.9+1.1 displays a radio continuum spectrum with a spectral index of 
$\alpha \approx -0.6$ \citep{land93,koth06}. In both cases, the steep spectra are explained 
by aging, although the recent discovery of a very young energetic pulsar in
G76.9+1.1 indicates a young age for the PWN \citep{arzo11}.
Since shell-type remnants are created by the shockwave of the supernova explosion they
always are limb-brightened. G141.2+5.0 is
well resolved in our data with a diameter of more than 4 beams. If it was a shell-type
SNR it would have an emission hole in the center or have an arc-like appearance.

Therefore we conclude that 
G141.2+5.0 is most likely a newly identified pulsar wind nebula.

\subsection{The distance of G141.2+5.0}
 
G141.2$+$5.0 is observed towards 
$b\simeq+$5$\degr$, well above the maximum latitude to which the Perseus arm is 
traced. Quadrant II Perseus arm \ion{H}{2} regions are found with a mean 
latitude $b=-$0$\fdg$2, with 34 out of 37 within $\left|b\right|<$2$\degr$ and 
none at $b\gtrsim$+3$\degr$ \citep[see catalogue of][]{fost13b}. Further, the 
systemic velocity of G141.2$+$5.0 is with $-53\pm 1$~km\,s$^{-1}$ significantly 
more negative than Perseus arm 
\ion{H}{1} ($v_{LSR}\simeq-$39~km~s$^{-1}$ in the midplane). Kinematically 
then, G141.2$+$5.0 is not within the Perseus arm, but in the Cygnus 
spiral arm, the next major star-forming arm about 2.5~kpc beyond Perseus up in 
the warped disc \citep[][]{fost10}. In \citet{fost13b} five nearby Cygnus arm 
\ion{H}{2} regions (Sh2-192, 193, 196, LBN\,676 \& Sh2-204) with comparable 
$\ell,~b,~v_{LSR}$ to G141.2$+$5.0 have a mean distance among them of 
3.8$\pm$1.1~kpc.

To estimate the distance to G141.2$+$5.0 we use the \ion{H}{1} modelling method
of \citet{fost06} to predict the line-of-sight run of velocity with 
heliocentric distance, that, along with a simple density model of the 
\ion{H}{1} disk, best describes the observed \ion{H}{1} profile in this 
direction. The variation in the fitted parameters among many modelling runs 
captures the uncertainty in the model and in the distance to the object. We 
also account for an uncertainty in the \ion{H}{1} shell itself of 
$\Delta v_{LSR}=\pm$3.4~km~s$^{-1}$ \citep[estimated from such shells around 
\ion{H}{2} regions; see][]{fost13b}. The final kinematic distance to 
G141.2$+$5.0 is 4.0$\pm$0.5~kpc.

G141.2$+$5.0 is likely the remnant of a massive star that originally formed in
the Cygnus arm, an active star-forming arm well populated with bright OB stars 
\citep{negu03} and \ion{H}{2} regions \citep{kime89}. Despite our favourable 
view 
no Cygnus arm supernova remnants (including PWNe) have been 
conclusively identified. G141.2$+$5.0 thus represents the first. 

\subsection{Some characteristics of the PWN G141.2+5.0}

The newly discovered PWN G141.2+5.0 has a diameter of about 4\,pc the same as 
the young PWN 3C\,58 \citep{koth13}. There is no published detection of a 
pulsar nearby, most likely because there has not been a deep survey for pulsars in this 
area of the Galaxy, yet. G141.2+5.0 adds to the growing number of PWNe that 
are found without an associated shell-type SNR. 
Since we still do not know why this is the case, in particular for young PWNe like the Crab nebula 
and 3C\,58, every new discovery will get us closer to finding the solution for 
this mystery. There are many possible explanations for the lack of a radio shell: low
explosion energy, low ambient density, or low ambient magnetic field, to name just
a few. Given the stark contrast between the number of
shell-type SNRs in the nearby Perseus arm and none identified in Cygnus, the
lack of an accompanying radio shell may not be a coincidence, but rather
reflective of a different nature of Cygnus arm supernovae and their 
environments. However, since we have only one example it may just be a coincidence.

G141.2+5.0 is the only member of this group that has a small HI bubble around it. The HI emission 
of this feature is, with a few Kelvin above the background, very faint and we only were able to 
detect it, because it is not confused by much other Galactic HI emission. The HI shell has a diameter
of about $12\arcmin$ which translates to 14~pc at a distance of 4~kpc. At this distance the shell 
contains about 35~M$_\odot$ ($\pm 50$~\%) of material, assuming that 10\,\% of the atoms
are helium and the HI gas is optically thin. The low brightness temperature makes this a valid assumption.
This would translate to a pre-bubble density of $n_0 \approx 0.8$~cm$^{-3}$. 


If we assume that the HI shell is the missing SNR shell, it would be close to merging 
with the interstellar medium (ISM)
as evident from the low expansion velocity and the fact that it is breaking apart. From
\citet{ciof88}, assuming solar abundances, we find for a highly evolved SNR:
\begin{equation}
\begin{split}
E_{0} & = 6.8\times 10^{43}\,\left [\frac{n_0}{\rm cm^{-3}}\right ]^{1.16}\,\left 
[\frac{v}{\rm km\,s^{-1}}\right ]^{1.35}\,\left [\frac{R}{\rm pc}\right ]^{3.16} \\
 & = 2.76\times 10^{47}~{\rm erg},
\end{split}
\end{equation}
here $E_{0}$ is the explosion energy, $n_0$ the ambient density, $v$ the
shock velocity, and $R$ is the radius. The dynamic age of the
shell would be about 350,000 years. This is an extremely low value for
the explosion energy and a very old SNR (and PWN).

If the HI shell is the stellar wind bubble of the progenitor star, the lack of a shell-type remnant 
could be explained
by assuming that the SNR shock wave did not reach the HI shell, yet, and is still expanding freely
into its low density interior. From \citet{weav77}, we find a dynamic age for this bubble of about
700,000 years. This is extremely young and only very few exceptionally massive stars have such a short
life. But those stars would produce a massive stellar wind depositing $10^{51}$~erg and a lot of
material into the ISM. However, stellar winds of massive stars have several phases, an early high 
velocity low-mass-loss wind during their main sequence time and a slow high-mass-loss wind later on.
\citet{koth07} found HI structures very similar to the shell we found around G141.2+5.0 in the massive
stellar cluster Westerlund~1. They suggested that these consist of recombined material lost by massive
stars in a late high-mass-loss wind. The supernova shockwave could have swept up that material into
this shell. Again we would need a highly evolved SNR and a low energy explosion.

Given the discussion above, we propose that G141.2+5.0 is a pulsar wind nebula and the
HI shell we found around it is the highly evolved shell-type supernova remnant consisting
of swept up high-mass-loss wind material ejected by the progenitor star. In this case
the progenitor must have been a massive star, probably of O-type.

G141.2+5.0 is only detected at radio frequencies. We did not find
a counterpart in publically available X-ray or $\gamma$-ray surveys such as
the Fermi Survey. This is not untypical for a highly evolved PWN 
\citep[e.g.][]{koth08b}. Pointed follow-up observations with the CHANDRA satellite
may provide crucial information in solving the mysteries around this newly discovered PWN.

\section{Conclusions}

G141.2+5.0 is a new radio pulsar wind nebula discovered in follow-up
observations of the Canadian Galactic Plane Survey, and the first radio 
PWN discovered since G63.7$+$1.1 was reported by \citet{wall97} some 17 years 
ago. The PWN is a resident of the Cygnus arm about 4~kpc away and is the first 
SNR discovered in that spiral arm. It displays a steep radio continuum spectrum and has a high level of linear
polarization. It sits inside an expanding HI shell, which very likely is related to the expanding shell-type remnant
of the same supernova explosion at a highly advanced phase of evolution. This 
may identify pulsar wind nebulae without an accompanying shell-type remnant as 
the result of a low energy supernova explosion.

\acknowledgments

The Dominion Radio Astrophysical Observatory is a National Facility
operated by the National Research Council Canada. TF thanks
the NRCC for support of his sabbatical year at DRAO. 
XS was supported by the Australian Research Council through grant FL100100114.
We thank P. Reich and T. Landecker for careful reading of the manuscript.

\bibliographystyle{aa}

\appendix


\begin{deluxetable}{rr@{$\pm$}ll}
   \tablewidth{0pc}
   \tablecolumns{4}
   \tablecaption{\label{tab:flux} Integrated flux densities $S$ at frequencies $\nu$ for the newly
   discovered PWN G141.2+5.0.
   All flux densities have been corrected for the contribution of the nearby point-like source
   (see text).}
   \tablehead{$\nu$ [MHz] & \multicolumn{2}{c}{$S$ [mJy]} & {Reference}}
   \startdata
 74 & 1100 & 240 & VLSSr \citep{lane12}\\
 151 & 490 & 160 & 7C(G) Survey \citep{vess98}\\
 232 & 355 & 80 & Miyun 232 MHz \citep{zhan97}\\
 327 & 345 & 100 & WENSS \citep{reng97}\\
 408 & 318 & 35 & This paper \\
1420 & 137 & 14 & NVSS \citep{cond98}\\
1420 & 120 & 8 & This Paper \\
4850 & 52 & 11 & SPECFIND \citet{voll10}\\
   \enddata
\end{deluxetable}

\end{document}